\documentclass[a4paper]{jpconf}
\usepackage{graphicx}
\begin{document}
\title{Phenomenology of the pseudoscalar glueball with a mass of 2.6 GeV}

\author{Walaa I. Eshraim and Stanislaus Janowski}

\address{Institute for Theoretical Physics, Goethe-University, Max-von-Laue-Str. 1, 
\\60438 Frankfurt am Main, Germany}

\ead{weshraim@th.physik.uni-frankfurt.de; janowski@th.physik.uni-frankfurt.de}

\begin{abstract}
We calculate the branching ratios of the lightest pseudoscalar glueball, $J^{PC}=0^{-+}$,
with a mass of 2.6 GeV predicted by lattice QCD by using a chiral Lagrangian.
This study is relevant in view of the planned PANDA
experiment at the upcoming FAIR facility, which will measure the
energy range in which the pseudoscalar glueball is predicted to
exist.
\end{abstract}
 
\section{Introduction}

Quantum Chromodynamics (QCD) is the theory of strongly interacting matter. In QCD one expects the existence of glueballs,
which are bound states made of two or more gluons, the gauge bosons of QCD, and are invariant under gauge transformations.
The glueballs are important objects in order to understand the nonperturbative behaviour of QCD
as well as the hadronic spectroscopy \cite{review}.

Calculations of lattice QCD in the pure Yang-Mills sector predict a complete
glueball spectrum \cite{Morningstar}, where the lightest glueball state is a scalar-isoscalar, $J^{PC}=0^{++}$,
and its mass is about 1.6 GeV. This energy region has been studied in a variety of
effective approaches, e.g. Refs. \cite{scalars,stani}. The second lightest glueball state has been
predicted to be a tensor state, $J^{PC}=2^{++}$, with a mass of 2.2 GeV, 
e.g. Ref. \cite{tensor} for a related phenomenological discussion. 
The third lightest glueball state predicted by lattice QCD 
is a pseudoscalar state, $J^{PC}=0^{-+}$, and has a mass of about 2.6 GeV.
This pseudoscalar glueball, denoted as $\tilde{G}\equiv gg$, is studied here: we calculate its decays
into three pseudoscalar mesons, $\tilde{G}\rightarrow PPP$, and a pseudoscalar and a scalar meson, $\tilde{G}\rightarrow PS$.
The interaction between the pseudoscalar glueball and the ordinary scalar and pseudoscalar mesons is described by the effective chiral
Lagrangian introduced in Refs. \cite{our,proc}. The branching ratios represent a parameter free prediction of our approach.

The PANDA experiment at the FAIR facility in Darmstadt is currently under construction and will use an 1.5 GeV antiproton beam
hitting a proton target at rest \cite{panda}. Therefore a center of mass energy higher than $\sim$2.5 GeV will be reached and the 2.6 GeV
pseudoscalar glueball can be directly produced as an intermediate state \cite{proc}.

\section{The chiral Lagrangian}

The effective chiral Lagrangian which couples the pseudoscalar
glueball $\tilde{G}\equiv gg$ with quantum
numbers $J^{PC}=0^{-+}$ to the ordinary scalar and pseudoscalar mesons reads
\cite{our,proc,schechter}:

\begin{equation}
\mathcal{L}_{\tilde{G}}^{int}=ic_{\tilde{G}\Phi}\tilde{G}\left(
{\textrm{det}}\Phi-{\textrm{det}}\Phi^{\dag}\right)\label{intlag},
\end{equation}
where $c_{\tilde{G}\Phi}$ is a coupling constant and $\Phi$ is a multiplet containing the ordinary scalar
and pseudoscalar mesons. In this study we consider three flavours, $N_{f}=3$, thus $c_{\tilde{G}\Phi}$ is dimensionless
and the multiplet $\Phi$ reads \cite{dick}:

\begin{equation}
\Phi=\frac{1}{\sqrt{2}}\left(
\begin{array}
[c]{ccc}%
\frac{(\sigma_{N}+a_{0}^{0})+i(\eta_{N}+\pi^{0})}{\sqrt{2}} & a_{0}^{+}%
+i\pi^{+} & K_{S}^{+}+iK^{+}\\
a_{0}^{-}+i\pi^{-} & \frac{(\sigma_{N}-a_{0}^{0})+i(\eta_{N}-\pi^{0})}%
{\sqrt{2}} & K_{S}^{0}+iK^{0}\\
K_{S}^{-}+iK^{-} & \bar{K}_{S}^{0}+i\bar{K}^{0} & \sigma_{S}+i\eta_{S}%
\end{array}
\right). \label{phimatex}%
\end{equation}

We first study the symmetries of the effective Lagrangian (\ref{intlag}).
The pseudoscalar glueball $\tilde{G}$ is made of gluons and is therefore chirally invariant.
The multiplet $\Phi$ transforms under the chiral symmetry as
$\Phi\rightarrow U_{L}\Phi U_{R}^{\dagger}$, where $U_{L(R)}
=e^{-i\theta_{L(R)}^{a}t^{a}}$ is an element of $U(3)_{R(L)}$.
Performing these transformations on the determinant of $\Phi$ it is easy to prove that this object is invariant under
$SU(3)_{L}\times SU(3)_{R}$, but not under the axial $U_{A}(1)$ transformation:
\begin{equation}
\mathrm{det}\Phi\rightarrow\mathrm{det}U_{A}\Phi U_{A}=e^{-i\theta_{A}%
^{0}\sqrt{2N_{f}}}\mathrm{det}\Phi\neq\mathrm{det}\Phi \,. \label{ua}
\end{equation}
This is in agreement with the chiral anomaly.
Consequently the effective Lagrangian (\ref{intlag}) possesses only the $SU(3)_{R}\times SU(3)_{L}$ symmetry. 
Further essential symmetries of the strong interacting matter are the parity $\mathcal{P}$ 
and charge conjugation $\mathcal{C}$. The pseudoscalar glueball and the multiplet $\Phi$ transform
under parity as
\begin{equation}
\tilde{G}(t,\vec{x})\rightarrow-\tilde{G}(t,-\vec{x}) \,, \
\Phi(t,\vec{x})\rightarrow\Phi^{\dagger}(t,-\vec{x})\label{p} \,, 
\end{equation}
and under charge conjugation as
\begin{equation}
\tilde{G}\rightarrow\tilde{G} \,, \ \Phi\rightarrow\Phi^{T}\label{c} \,.
\end{equation}
Performing these discrete transformations $\mathcal{P}$ 
and $\mathcal{C}$ on the effective Lagrangian (\ref{intlag}) leave it unchanged. In conclusion, one can say that the
symmetries of the effective Lagrangian (\ref{intlag}) are in agreement with the symmetries of the QCD Lagrangian.

Let us now consider the assignment of the ordinary mesonic d.o.f. in Eq. (\ref{intlag}) or (\ref{phimatex}).
In the scalar sector the field
$\vec{a}_{0}$ corresponds to the physical isotriplet state $a_{0}(1450)$ and
the scalar kaon field $K_{S}$ to the physical isodoublet state
$K_{0}^{\star}(1430)$ \cite{PDG}. The fields
$\sigma_{N}\equiv(\bar{u}u+\bar{d}d)
/\sqrt{2}$ and $\sigma_{S}\equiv \bar{s}s$ are the bare nonstrange and strange isoscalars and 
they correspond to the physical isoscalars $f_{0}(1370)$ and $f_{0}(1710)$, respectively \cite{dick}. Scalars below 1 GeV
are predominantly tetraquarks or mesonic molecular states, see Refs. \cite{tetraquark,lowscalars} and references therein.
As shown in Ref. \cite{dick} the
mixing of the bare fields $\sigma_{N}$ and $\sigma_{S}$ is small and is neglected at this stage.
In the pseudoscalar sector we assign the fields $\vec{\pi}$ and $K$ to the physical pion isotriplet
and the kaon isodoublet \cite{PDG}. The bare quark-antiquark fields $\eta_{N}%
\equiv (\bar{u}u+\bar{d}d) /\sqrt{2}$ and $\eta_{S} \equiv\ \bar{s}s$
are the nonstrange and strange  mixing contributions of the physical states $\eta$ and $\eta^{\prime}$ \cite{dick}.
In the effective Lagrangian (\ref{intlag}) there exist a mixing between the bare pseudoscalar glueball $\tilde{G}$
and the both bare fields $\eta_{N}$ and  $\eta_{S}$, but, due to the large mass difference between the pseudoscalar glueball
and the pseudoscalar quark-antiquark fields, it turns out that its mixing is very small and is therefore negligible.

In accordance with the spontaneous breaking of the chiral symmetry we shift the scalar-isoscalar
fields by theirs vacuum expectation values $\sigma_{N}\rightarrow\sigma_{N}+\phi_{N}$ and
$\sigma_{S}\rightarrow \sigma_{S}+\phi_{S} \label{shift}$, where $\phi_{N}$ and
$\phi_{S}$ are the corresponding chiral condensates. In order to be consistent with the full
effective chiral Lagrangian of the extended Linear Sigma Model \cite{dick,Susanna,Paper1}
we have to consider the shifting of the axialvector fields and thus to redefine the wave function of the pseudoscalar fields
\begin{equation}
\vec{\pi}\rightarrow Z_{\pi}\vec{\pi} \,, \ K\rightarrow Z_{K}%
K \,, \ \eta_{N,S}\rightarrow Z_{\eta_{N,S}}\eta_{N,S}\;, \label{psz}%
\end{equation}
where $Z_i$ are the renormalization constants of the corresponding wave functions \cite{dick}.

\section{Results}

The unknown coupling constant $c_{\tilde{G}\Phi}$ of the effective chiral Lagrangian (\ref{intlag}) is the only
free parameter of our approach. In order to make parameter free predictions regarding the future PANDA experiment
at the FAIR facility we calculated the branching ratios of
the two- and three-body decays of the pseudoscalar glueball $\tilde{G}$.
When evaluating the decay widths from Eq. (\ref{intlag}), we used, in agreement with lattice QCD,
the mass of the pseudoscalar glueball $M_{\tilde{G}}=2.6$ 
GeV \cite{Morningstar,our,proc}.
Our results for the decay channels $\tilde{G}\rightarrow PPP$ and $\tilde{G}\rightarrow PS$ are presented in Table 1 and Table 2
respectively, where $\Gamma_{\tilde{G}}^{tot}=\Gamma_{\tilde{G}\rightarrow
PPP}+\Gamma_{\tilde {G}\rightarrow PS}$ is the total decay width, \cite{our,proc}.

\begin{center}%
%TCIMACRO{\TeXButton{B}{\begin{table}[h] \centering}}%
%BeginExpansion
\begin{table}[h] \centering
%EndExpansion%
\begin{tabular}
[c]{|c|c|}\hline Quantity &  Value \\\hline
$\Gamma_{\tilde{G}\rightarrow KK\eta}/\Gamma_{\tilde{G}}^{tot}$ &
$0.049$ \\\hline $\Gamma_{\tilde{G}\rightarrow
KK\eta^{\prime}}/\Gamma_{\tilde{G}}^{tot}$ & $0.019$ \\\hline
$\Gamma_{\tilde{G}\rightarrow\eta\eta\eta}/\Gamma_{\tilde{G}}^{tot}$
& $0.016$ \\\hline
$\Gamma_{\tilde{G}\rightarrow\eta\eta\eta^{\prime}}/\Gamma_{\tilde{G}}^{tot}$
& $0.0017$ \\\hline
$\Gamma_{\tilde{G}\rightarrow\eta\eta^{\prime}\eta^{\prime}}/\Gamma_{\tilde
{G}}^{tot}$ & $0.00013$ \\\hline $\Gamma_{\tilde{G}\rightarrow
KK\pi}/\Gamma_{\tilde{G}}^{tot}$ & $0.46$ \\\hline
$\Gamma_{\tilde{G}\rightarrow\eta\pi\pi}/\Gamma_{\tilde{G}}^{tot}$
& $0.16$ \\\hline
$\Gamma_{\tilde{G}\rightarrow\eta^{\prime}\pi\pi}/\Gamma_{\tilde{G}}^{tot}$
& $0.094$ \\\hline
\end{tabular}%
%TCIMACRO{\TeXButton{Caption}{\caption
%{Branching ratios for the decay of the pseudoscalar glueball $\tilde
%{G}$ into three pseudoscalar mesons.}}}%
%BeginExpansion
\caption{Branching ratios for the decay of the pseudoscalar
glueball $\tilde
{G}$ into three pseudoscalar mesons.}%
%EndExpansion%
%TCIMACRO{\TeXButton{E}{\end{table}}}%
%BeginExpansion
\end{table}%
%EndExpansion

\end{center}

\begin{center}%
%TCIMACRO{\TeXButton{B}{\begin{table}[h] \centering}}%
%BeginExpansion
\begin{table}[h] \centering
%EndExpansion%
\begin{tabular}
[c]{|c|c|}\hline Quantity & Value \\\hline
$\Gamma_{\tilde{G}\rightarrow KK_{S}}/\Gamma_{\tilde{G}}^{tot}$ &
$0.059$ \\\hline $\Gamma_{\tilde{G}\rightarrow
a_{0}\pi}/\Gamma_{\tilde{G}}^{tot}$ & $0.083$ \\\hline
$\Gamma_{\tilde{G}\rightarrow\eta\sigma_{N}}/\Gamma_{\tilde{G}}^{tot}$
& $0.028$\\\hline
$\Gamma_{\tilde{G}\rightarrow\eta\sigma_{S}}/\Gamma_{\tilde{G}}^{tot}$
& $0.012$\\\hline
$\Gamma_{\tilde{G}\rightarrow\eta^{\prime}\sigma_{N}}/\Gamma_{\tilde{G}}%
^{tot}$ & $0.019$ \\\hline
\end{tabular}%
%TCIMACRO{\TeXButton{Caption}{\caption
%{Branching ratios for the decay of the pseudoscalar glueball $\tilde
%{G}$ into a scalar and a pseudoscalar meson.}}}%
%BeginExpansion
\caption{Branching ratios for the decay of the pseudoscalar
glueball $\tilde
{G}$ into a scalar and a pseudoscalar meson.}%
%EndExpansion%
%TCIMACRO{\TeXButton{E}{\end{table}}}%
%BeginExpansion
\end{table}%
%EndExpansion

\end{center}

The largest contribution to the total decay width is given by the following decay channels:
$KK\pi$, which contributes to almost 50\%, as well as $\eta\pi\pi$ and $\eta'\pi\pi$, where each one contributes of about 10\%.
Although the decays of the pseudoscalar glueball into the scalars-isoscalars contribute only 6\% to the total decay width, they 
should be treated carefully because they mix with each other and its assignment to the physical states is up to now unclear.
In order to make more precisely statements in the scalar-isoscalar sector detailed calculations  are necessary and are presently ongoing.
A further interesting outcome of our approach is that the decay of the pseudoscalar glueball into three pions, $\tilde{G}\rightarrow\pi\pi\pi$,
is not allowed.

\section{Conclusions}

We have presented a chirally invariant three-flavour
effective Lagrangian with scalar and pseudoscalar quark-antiquark states and a
pseudoscalar glueball.
We have calculated two- and three-body decay processes of the
pseudoscalar glueball with a mass of 2.6 GeV, as evaluated by lattice QCD.
We predict that the decay channel
$\tilde{G}\rightarrow KK\pi$ is predominant and $\tilde{G}\rightarrow \eta\pi\pi$ as well as $\tilde{G}\rightarrow \eta'\pi\pi$
are the next dominant ones \cite{our,proc}. Moreover, the decay channel $\tilde{G}\rightarrow\pi\pi\pi$ is not allowed.
\\
The results presented in this work can be tested in the upcoming PANDA experiment at the
FAIR facility \cite{panda}.

\section*{Acknowledgments}

The authors thank Francesco Giacosa and Dirk H. Rischke for useful
discussions. W.E.\ acknowledges support from DAAD and HGS-HIRe,
S.J.\ acknowledges support from H-QM and HGS-HIRe.

\end{document}